\documentclass{PoS}
\usepackage{amsmath,amssymb}
\usepackage{graphics}

\newcommand{\bra}[1]{\left\langle #1 \right|}
\newcommand{\ket}[1]{\left| #1 \right\rangle}
\newcommand{\braket}[2]{\langle #1| #2 \rangle}

\title{Dynamical overlap fermions with increased topological tunnelling}

\ShortTitle{Dynamical overlap fermions with increased topological tunnelling}

\author{\speaker{Nigel Cundy}%
        \\
        University of Regensburg, Universit\"atstrasse 31, 93040 Regensburg, 
Germany.\\
        E-mail: \email{nigel.Cundy@physik.uni-regensburg.de}}

\author{Stefan Krieg\\
J\"ulich Supercomputing Centre,
Forschungszentrum J\"ulich GmbH,
52425 J\"ulich,
Germany.\\
        E-mail: \email{s.krieg@fz-juelich.de }}

\author{Thomas Lippert\\
      J\"ulich Supercomputing Centre,
Forschungszentrum J\"ulich GmbH,
52425 J\"ulich,
Germany.\\
        E-mail: \email{th.lippert@fz-juelich.de}}

\author{Andreas Sch\"afer\\
         University of Regensburg, Universit\"atstrasse 31, 93040 Regensburg, 
Germany.\\
        E-mail: \email{andreas.schaefer@physik.uni-regensburg.de}}

\abstract{We present two improvements to our previous dynamical overlap HMC algorithm. We introduce a new method of differentiating the eigenvectors of the Kernel operator, which removes an instability in the fermionic force. Secondly, by simulating part of the fermion determinant exactly, without pseudo-fermions, we are able to increase the rate of topological tunnelling by a factor of more than ten, reducing the auto-correlation. }

\FullConference{The XXV International Symposium on Lattice Field Theory\\
                 July 30 - August 4 2007\\
                 Regensburg, Germany}

\begin{document}

\section{Introduction}\label{sec:1}
The overlap operator ~\cite{Narayanan:1993ss} is the only known lattice Dirac operator with an exact lattice chiral symmetry~\cite{Luscher:1998pqa}. Since chiral symmetry is important for many low energy observables, it is desirable to use overlap fermions. However, using a full QCD simulation with overlap fermions presents a number of algorithmic challenges. In this paper, we address, and present solutions for, two outstanding algorithmic issues. 

The problem of the Dirac $\delta$-function in the fermionic force when changing topological charge has been resolved by using a transmission/reflection algorithm, similar to the case of a classical mechanics particle approaching a potential wall. The original formulation ~\cite{Fodor:2003bh,Cundy:2005pi} has subsequently been improved in ~\cite{Cundy:2005mr} to maximise the rate of topological charge change for a given action jump. Our method is described in these references.

To differentiate the overlap operator in the Hybrid Monte Carlo (HMC) molecular dynamics (MD), it is necessary to differentiate the eigenvectors and eigenvalues of a sparse matrix. Previous methods have led to instabilities when there are degenerate eigenvalues. In section \ref{sec:2} we discuss a new method which avoids these instabilities~\cite{Cundy:2007df}. 

The topological auto-correlation depends on the rate of topological activity. To have high rate of topological index change, it is necessary to reduce the discontinuity in the action at this point.  In section \ref{sec:3}, we outline a new method for reducing this action jump~\cite{cundyforthcoming}.

Our overlap operator is defined as
\begin{gather}
D = (1+\mu) + \gamma_5 (1-\mu) \text{sign}(Q).
\end{gather}
In our tests, $Q$ is the standard Wilson operator, 
\begin{gather}
Q = \gamma_5 \left[\delta_{xy}-\kappa\left( (1-\gamma_{\mu})U_{\mu}(x)\delta_{y,x+\mu} + (1+\gamma_{\mu})U^{\dagger}_{\mu}(x-\mu)\delta_{y,x-\mu}\right)\right],
\end{gather}
 with $\kappa = 0.2$ and, in section \ref{sec:3}, two levels of stout smearing~\cite{Morningstar:2003gk} at parameter $0.1$. Our numerical tests are performed on $8^316$ lattices at a lattice spacing of about $0.15$fm (measured using $r_0$), with quark masses $\mu = 0.03,0.04,0.05$, corresponding to pion masses in the range $500-1000$MeV. 
\section{Eigenvector mixing}\label{sec:2}
\subsection{Differentiating Eigenvectors}

The eigenvalues $\lambda_i$ and eigenvectors $|\psi_i\rangle$ of a matrix $Q$ are defined by
\begin{gather}
Q |\psi_i\rangle = \lambda_i |\psi_i\rangle.
\end{gather}
After a small change to the matrix, $\delta Q$, the new eigenvalue equation is
\begin{gather}
(Q+\delta Q) |\psi'_i\rangle = \lambda'_i|\psi'_i\rangle.
\end{gather}
We can expand the new eigenvectors in terms of the old basis
\begin{gather}
 |\psi'_i\rangle = |\psi_i\rangle + \sum_j (\cos\theta_{ij} - 1) |\psi_i\rangle + e^{i\phi_{ij}}\sin\theta_{ij} |\psi_j\rangle.
\end{gather}
We assume that only one of the mixing angles $\theta_{ij}$ is large, so it is unimportant that $\ket{\psi'_i}$ is not normalised\footnote{A more general expression can easily be constructed should this assumption break down.}. The mixing angles $\theta_{ij}$ and $\phi_{ij}$ are
\begin{align}
\tan2\theta_{ij} =& \frac{2\sqrt{\langle\psi_i|\delta Q|\psi_j \rangle\bra{\psi_j} \delta Q \ket{\psi_i}}}{\lambda_j - \lambda_i + \langle\psi_j|\delta Q|\psi_j\rangle - \langle\psi_i|\delta Q|\psi_i\rangle}\nonumber\\
e^{i\phi_{ij}} =& \sqrt{\frac{\bra{\psi_j}\delta Q\ket{\psi_i}}{\bra{\psi_i}\delta Q\ket{\psi_j}}}.\label{eq:tantheta}
\end{align}
Expanding $\theta$ and $\phi$ in $\tau/(\lambda_i - \lambda_j)$ gives to lowest order 
\begin{align}
\delta |\psi_i\rangle =& \sum_{j\neq i} |\psi_j\rangle \frac{\bra{\psi_j}\delta Q \ket{\psi_i}}{\lambda_j - \lambda_i} = \frac{1}{Q-\lambda_i}(1-|\psi_i\rangle\langle\psi_i|)\delta Q |\psi_i\rangle,\label{eq:badforce}
\end{align}
which agrees with other methods (see, for example, ~\cite{Cundy:2005mr}). Equation (\ref{eq:badforce}) breaks down when $\lambda_i-\lambda_j$ is small. In this situation, it is necessary to use the exact expressions for the mixing angles (\ref{eq:tantheta}).
Defining $T^n_{\mu}(x)$ as the eight SU($3$) generators on one link, we can write \begin{gather}\delta Q_{ij} = \langle\psi_i|\delta Q|\psi_j\rangle = \tau\pi^n_{\mu}(x) \alpha_{ij}^{n,x,\mu},\end{gather} where
 $\pi$ represents the MD momentum, and the vectors $\alpha_{ij}$ are given by
\begin{gather}
\alpha_{ij}^{n,x,\mu} = -i\kappa\langle\psi_i|\gamma_5\left[(1-\gamma_\mu)T^n_{\mu}(x) U_{\mu}(x)\delta_{y,x+\mu} - (1+\gamma_{\mu}) U^{\dagger}_{\mu}(x)T^i_{\mu}(x)\right]\delta_{y,x-\mu}|\psi_j\rangle.
\end{gather}
The generalisation to the smeared operator is trivial~\cite{Cundy:2007df}. The NAC force, $F^{NAC}$, constructed for the overlap operator in ~\cite{Cundy:2007df}, is proportional to $\alpha_{ij}$, and a function of only $\pi\alpha$ and the gauge field, $U$.
We can construct a reversible algorithm by combining forward and backward half-steps (using a quick and reversible iterative procedure for the backward step). The force is not area conserving, but it is possible to use a non-area conserving force by calculating the Jacobian, $J$, and including $\log J$ in the HMC accept/reject step (see ~\cite{Cundy:2005mr} for an example). $J$ can be written as
\begin{gather}
J = \left|1 + \alpha_{ij}^{n,x,\mu} \frac{\partial F_{ij}}{\partial \pi^{m,y,\nu}}\right| = \left|1+ A_{ij,kl} \alpha_{ij}\alpha_{kl}\right|.
\end{gather}
It is now a simple task to rewrite $\alpha_{ij}$ in terms of an orthonormal complete basis $\alpha'$, and use  
\begin{gather}
\det\left[1+A'\alpha'{\alpha'}^{\dagger}\right]=\det\left[1+A'\right]
\end{gather}
to calculate $J$. $\log J$ scales with $\tau^3$, and does not affect the HMC acceptance rate.


\subsection{Numerical results}

\begin{figure}
\begin{center}
\begin{tabular}{c}
\includegraphics[height = 8cm]{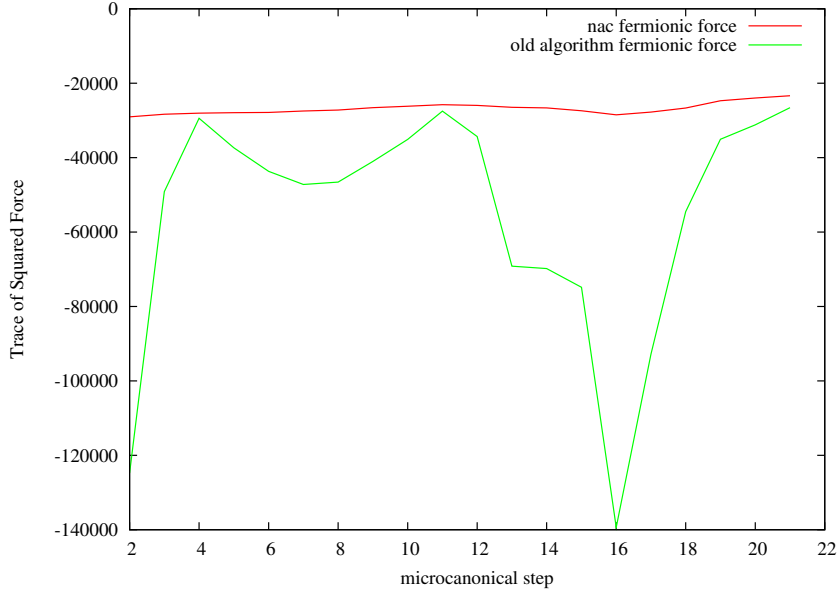}\\
\end{tabular}
\end{center}
\caption{Comparison of the trace of the square of the fermionic forces for the proposed and old algorithms with $\tau = 0.016$ on one of the $\mu = 0.05$ trajectories.}\label{fig:comparison}
\end{figure}

In figure \ref{fig:comparison}, we compare the fermionic forces for the old algorithm and the NAC algorithm across a typical HMC trajectory (using overlap fermions with no smearing). It is clear that the old method gives an unstable force, which cannot be used, while the NAC force is stable. In all our test trajectories, we have not observed a $\log J$ larger than 0.3.
\section{Determinant factorisation}\label{sec:3}
\subsection{Introduction}
The topological auto-correlation is related to the rate of topological activity. We need to be able to measure this rate, from which we can estimate the auto-correlation. Since instanton anti-instanton pairs are difficult to observe, our best way of measuring the auto-correlation is from topological charge changes. The probability of transmission scales as $\min(1,e^{\Delta S})$~\cite{Cundy:2005mr}, where $\Delta S$, the action discontinuity at the topological sector boundary, scales as $\mu^{-2}$. Therefore it is necessary to reduce $\Delta S$. 

However, the determinant of the actual Dirac operator, $\log\det(DD^{\dagger})$, scales as $\log\mu$ rather than $\mu^{-2}$~\cite{Egri:2005cx}. The difference is caused the different functional forms of the pseudo-fermion estimate and the actual determinant. It has already been shown that adding additional pseudo-fermions reduces $\Delta S$ considerably~\cite{Hasenbusch:2001ne,Degrand:2004nq}, and this (with one additional pseudo-fermion) is our algorithm A, which we test against our new methods. We propose~\cite{cundyforthcoming} factorising the determinant into a large continuous part, which can be treated with pseudo-fermions, and a discontinuous determinant small enough to be calculated exactly. 
\subsection{Algorithm C} 
Algorithm C factorises the determinant using
\begin{align}
\det\left[\gamma_5\frac{1+\mu}{1-\mu} + \epsilon(Q)\right] =& \det\left[\gamma_5\frac{1+\mu}{1-\mu} + \tilde{\epsilon}(Q)\right]\det\left[\delta_{ij} + \langle\psi_i|\frac{1}{\gamma_5\frac{1+\mu}{1-\mu} + \tilde{\epsilon}(Q)}|\psi_j\rangle(\epsilon(\lambda_j) - \tilde{\epsilon}(\lambda_j)\right]\nonumber\\
=&\det[D_1] \det[D_2].
\end{align}
$\tilde{\epsilon}$ is a continuous approximation of the sign function, only differing for the smallest $n$ eigenvalues below a cutoff $\Lambda$. In our tests, we used a Zolotarev rational approximation. We simulate $\det[D_1]$ using pseudo-fermions; $\det[D_2]$ is calculated by standard methods. We add $-\log\det[D_2]$ to the HMC action.
 To maintain a high HMC acceptance we calculate $\log\det[D_2]$'s force (unlike the proposal in~\cite{Schaefer:2006bk}). Differentiating $\log\det[D_2]$ is straight-forward using the methods of section \ref{sec:2}. 
Algorithm C requires an inversion of $D_1$ for each eigenvalue projected. While deflation methods~\cite{Cundy:2005mn} reduce this cost, it could lead to difficulties on larger volumes.\footnote{We cannot decrease $\Lambda$ without increasing the force; therefore the number of projected eigenvalues will increase as the volume increases.} 

\subsection{Algorithm F} 

To avoid this additional cost, algorithm F projects out just one vector, $\ket{a}$, which is equal to the smallest eigenvector at the moment of crossing. Thus, we write
\begin{align}
\det[1 + \gamma_5\epsilon(Q)] = &\det\left[1+\gamma_5\epsilon(Q)(1-\ket{a}\bra{a})\right]\det\left[1 + \frac{1}{1+\gamma_5\epsilon(Q)(1-\ket{a}\bra{a}])}\gamma_5\epsilon(Q)\ket{a}\bra{a}\right]\nonumber\\
=&\det[D_3]\det[D_4]
\end{align} 
To ensure $D_3$ is both continuous and practical, $\ket{a}$ must satisfy $\ket{a} = \ket{\psi_i}$ at $\lambda_i = 0$ and $\braket{\psi_i}{a} = 0$ for $\lambda_i^2 > \Lambda^2$ (where $\Lambda^2$ is some suitable eigenvalue cut-off). For the algorithm to remain stable at larger lattice volumes $\ket{a}$ and its differential must be continuous, $d/d\lambda_i\ket{a}$ must be sufficiently small, the eigenvalues and eigenvectors must be differentiated using the results of section \ref{sec:2} and the approximate overlap operator $D_3$ must not have any exceptional configurations. We construct $\ket{a}$ from $n$ eigenvectors using
\begin{gather}
\ket{a} = \beta_0\ket{\psi_0}\frac{\braket{\psi_0}{\Gamma}}{|\braket{\psi_0}{\Gamma}|} + \beta_1\ket{\psi_1}\frac{\braket{\psi_1}{\Gamma}}{|\braket{\psi_1}{\Gamma}|} + \ldots,
\end{gather}
where $\Gamma$ is a constant vector used to fix the relative phase of the eigenvectors, and currently we use
\begin{gather}
\tan\frac{\pi}{2}\beta_{i} = \lambda_{i}^4\Lambda^{4(n-1)}/(\Lambda^2 - \lambda_{i}^2)^2\prod_{j\neq i}\lambda_j^4,
\end{gather}
although we are still searching for the best function to use for $\beta$. Algorithm C has shown stable MD with high HMC acceptance and good reversibility on our test $8^316$ ensembles.\footnote{A variant of algorithm C, not using our method of differentiating the eigenvectors and not treating the small matrix in the molecular dynamics, has encountered problems on $12^324$ volumes~\cite{DegrandSchaeferPersonalCommunication}. We suspect that this was because the range of the approximate sign function was set to small, and a less efficient method was used when differentiating eigenvectors.} Algorithm F has encountered large differentials of $\beta$ with respect to the eigenvalues, which may cause problems on larger volumes, although the HMC acceptance rate is still acceptable on our lattices. 
 We do not expect any scaling of the action jump with the lattice volume.
\subsection{Numerical results}\label{results:3}
\begin{figure}
\begin{tabular}{c c}
\includegraphics[width=7.5cm]{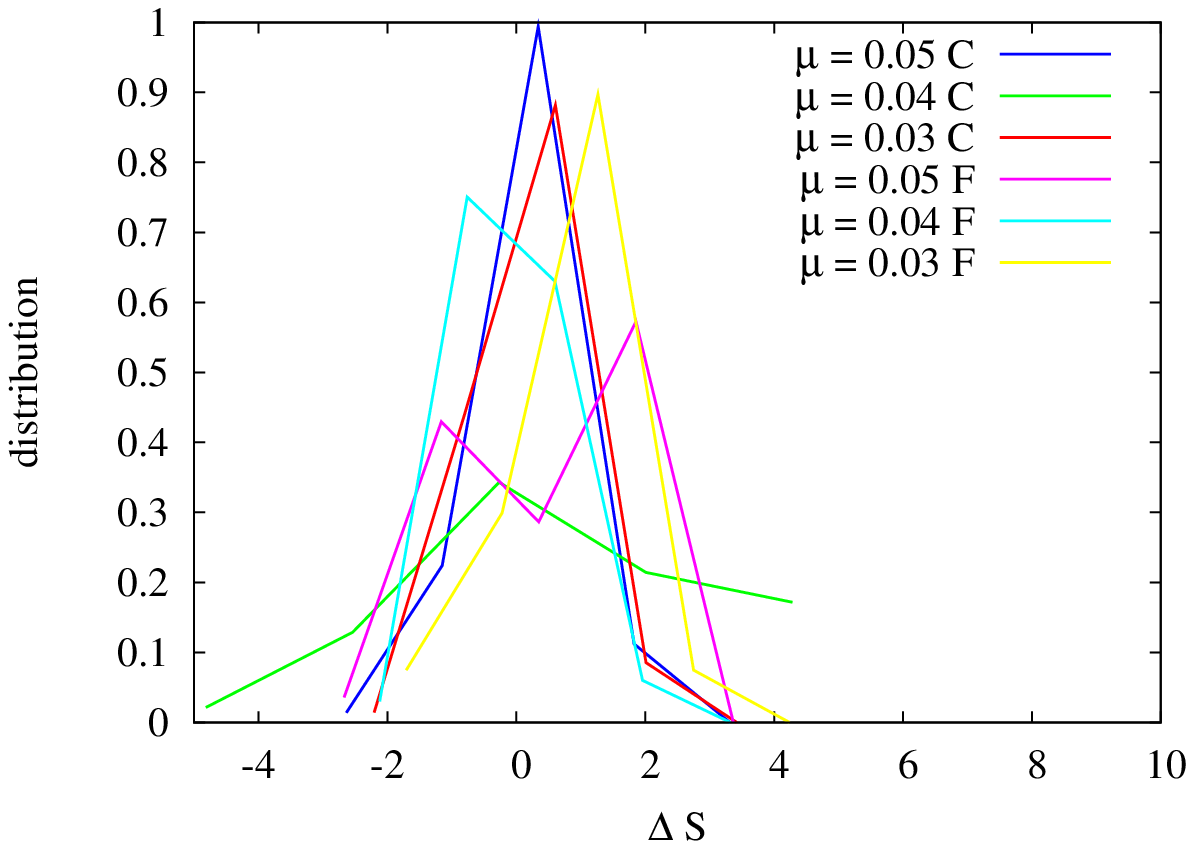}&
\includegraphics[width=7.5cm]{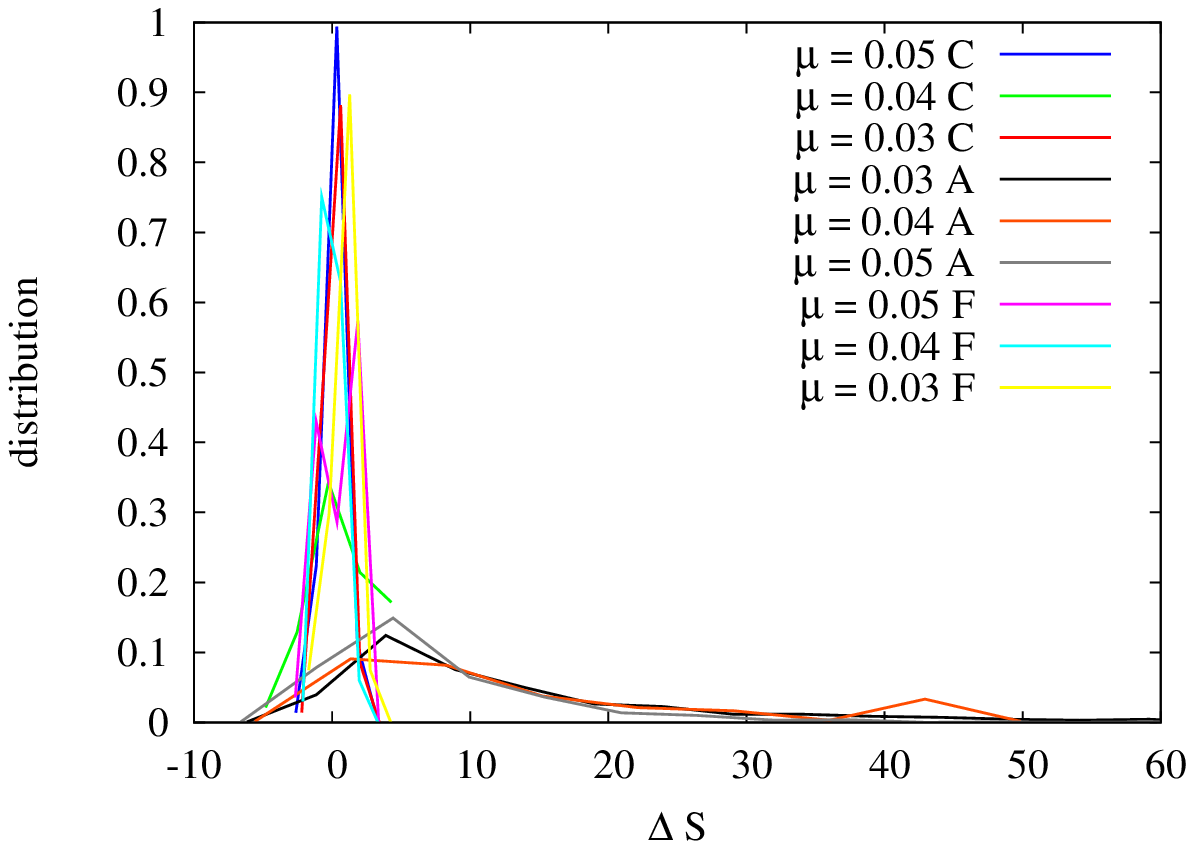}
\end{tabular}
\caption{The distribution of the action discontinuity $\Delta S$ for algorithms C and F (left), and A, C and F (right), at masses $\mu = 0.03,0.04$ and $0.05$.}\label{fig:1}
\end{figure}
\begin{table}
\begin{center}
\begin{tabular}{l|l l l| l l l}
$\mu$&A&C&F&A&C&F\\
\hline
0.03&14.0(7)&0.28(8)&{0.70(21)}
&21.9 &0.78&1.13
\\
0.04&13.8(10)&{0.48(32)}&{-0.06(11)}
& 19.5&2.4&0.86
\\
0.05&7.8(4)&0.23(7)&{0.22(22)}
& 13.6&0.72&1.3
\end{tabular}
\end{center}
\caption{The mean values of $\Delta S$ for algorithms A,C and F at three quark masses $\mu$ (left) and the standard deviations (right).}
\label{tab:1}
\end{table}

Figure \ref{fig:1} and table \ref{tab:1} give preliminary results for the action jump on our $8^316$ test configurations. $\Delta S$ is much smaller for algorithms C and F than for algorithm A,
 and the distribution is much narrower. The results for both algorithms C and F give a high transmission rate for all $\mu$. We see little variation with the quark mass with these methods, unlike algorithm A. This suggests transmission may be possible with much smaller $\mu$.
\section{Conclusion}\label{sec:4}
We have developed two additions to the overlap HMC algorithm designed to significantly improve simulations at large volume and small fermion mass. By employing a non area conserving algorithm to differentiate the eigenvectors and eigenvalues of the Dirac operator, we remove various instabilities that otherwise are encountered in the fermionic force. By factorising the fermion determinant, we decrease the action jump at the topological sector boundary considerably (we have observed transmission rates of up to 80\%), reducing the auto-correlation by an order of magnitude. Thus, with dynamical overlap fermions, it is possible to accurately sample all topological sectors efficiently. It remains an open question whether other lattice formulations will, at small lattice spacing, suffer from the large auto-correlations which we have now avoided.  
\section*{Acknowledgements}
Numerical simulations were run on the Cray-XD1 and Blue-Gene/L at the J\"ulich Supercompuing Center, Forschungszentrum J\"ulich. We thank Tom Degrand, Stefan Sch\"afer and Tony Kennedy for useful discussions. This work was supported by the DFG (FOR 465)

\bibliographystyle{pos}
\bibliography{proceedings}

\end{document}